# Core Electron Heating By Triggered And Ordinary Ion Acoustic Waves In The Solar Wind

By F.S. Mozer[1,2], S.D. Bale[1,2], C.A. Cattell[3], J. Halekas[4], I.Y. Vasko[1], J.L. Verniero[5] and P.J. Kellogg[3]
1. Space Sciences Laboratory, University of California, Berkeley, USA
2. Physics Department, University of California, Berkeley, USA
3. University of Minnesota, Minneapolis, Mn, USA
4. University of Iowa, Iowa City, Iowa, USA
5. Goddard Space Flight Center, Greenbelt, Md, USA

Orbits six through nine of the Parker Solar Probe have been studied to show that solar wind core electrons emerged from 15 solar radii with temperatures that were constant to within ~10% although the solar wind speed varied from 300 to 800 km/sec. After leaving 15 solar radii, the core electrons were isotropically heated as much as a factor of two below 30 solar radii by triggered and natural ion acoustic waves. To distinguish this wave heating from processes occurring inside 15 solar radii, the electron temperature is modeled as having two components, the base temperature observed at 15 solar radii and the temperature resulting from wave interactions between 15 solar radii and the observing point. As further justification of this temperature model, there were about a dozen intervals during the four orbits, summing to about 250 hours of total time, during which there were few waves, $T_E/T_I$ was small, and the core electron temperature was close to or slightly greater than the base temperature. Both this base temperature and the temperature resulting from wave heating decreased with radius due to the adiabatic expansion of the solar wind. The waves that produced the electron core heating were triggered [Mozer et al, 2021] or normal {Mozer et al, 2020A] ion acoustic waves. They are the dominant wave modes at frequencies greater than 100 Hz at solar distances between 15 and 30 solar radii.

**Introduction**

Ion acoustic waves have been observed by many satellites in the solar wind [Gurnett and Anderson, 1977; Gurnett and Frank, 1978; Kurth et al, 1979; Lin et al, 2001; Mozer et al, 2020A; Pisa et al, 2021]. They are wideband, short duration waves which differ greatly from the triggered ion acoustic waves (TIAW) seen at 20-30 solar radii on the Parker Solar Probe (PSP) [Mozer et al, 2021], in that these latter waves are narrowband, long duration (hours to days) waves that often appear as shock-like pulses at rates of a few Hz. The purpose of this paper is to study the effects of these ion acoustic waves on the electron plasma. It has long been known that the core solar wind electrons must be heated as they move away from the Sun [Hartle and Sturrock, 1968] to overcome some of the temperature loss associated with their adiabatic expansion, but the heating mechanism has not previously been identified in experimental data. The role of ion acoustic waves in this heating has been discussed theoretically [Dum, 1978] and via calculations [Kellogg, 2020]. This paper will further examine their role as a mechanism for electron heating, thereby addressing the PSP mission science goal, to "trace the flow of energy that heats…the solar corona and solar wind" [Fox, 2016]. The Fields [Bale et al, 2016; Malaspina, 2016; Mozer et al, 2020B] and SWEAP [Halekas et al, 2020; Kasper et al, 2016; Whittlesey et al, 2020] instruments on the Parker Solar Probe obtained the data presented in this paper.

To distinguish between the base temperature at 15 solar radii and the temperature increase resulting from wave heating between 15 and 30 solar radii, the PSP data are analyzed in terms of a model for the electron core parallel or perpendicular temperature versus radius in which

$$T_E = T_B + T_W$$

where
- $T_E$ is the electron temperature at the measurement point
- $T_B$ is the base electron temperature observed at 15 solar radii
- $T_W$ is the temperature increase achieved by wave-particle heating at all locations between 15 solar radii and the measurement point.

As will be suggested by the Parker Solar Probe data for orbits 6, 7, 8, and 9, the base electron temperature was approximately independent of source conditions other than that it weakly anti-correlated with the solar wind speed.

Both $T_B$ and $T_W$ decrease with solar radius because of cooling associated with the adiabatic expansion of the solar wind. To derive an expression for this cooling, it is assumed that $pV^\gamma$ is constant, where $P = nkT_E$ is the plasma pressure, n is the density, V is the gas volume and $\gamma = 5/3$ for a monatomic gas. Since $n \propto R^{-2}$ and $V \propto R^2$, where R is the distance from the Sun in units of solar radii, these equations combine to yield $T_B = 1900 R^{-4/3}$ for expansion of the electron gas without heating, where $T_B$ is the temperature in eV and the constant is determined by calibration of the PSP-measured $T_B$ versus radius. It will be found that $T_B$ is approximately constant for orbits 6, 7, 8, and 9, in spite of the different plasma conditions of the different orbits and the fact that the spacecraft crossed large regions of solar longitude near each perigee.

**Data**

Figure 1 presents data from the orbit 6 and orbit 8 perihelia. Panels 1A and 1G give the electric field spectra, panels 1B and 1H give $T_E$ (in black) and $T_B$ (in red), panels 1C and 1I give the ion perpendicular temperature, panels 1D and 1J give the solar wind speed, panels 1E and 1K give $T_E/T_I$, and panels 1F and 1L give the spacecraft distance from the Sun in units of solar radii. Halekas et al {2020} provided the electron temperature data shown throughout this paper. As seen in panel 1A, there were two types of wave activity, the broadband short duration normal ion acoustic waves that appear as vertical lines and the narrow band, longer duration waves at the beginning of the interval and that are called triggered ion acoustic waves. Other than during the first ~15% of panel 1B, the electron temperature followed the red $T_B$ curve with ~10% excursions that anti-correlated with variations of $T_I$ and the solar wind speed of panels 1C and 1D during times when the solar wind speed varied from 300 to 800 km/sec. Thus, the base electron temperature only weakly depended on any properties inside of 15 solar radii for about 75 hours. In panel 1E, $T_E/T_I$ was ~1 during the passage after the first ~15% of the figure, which suggests that ion acoustic wave growth was small due to Landau damping.

During the first ~15% of panel 1B, the electron temperature was significantly greater than $T_B$, which shows that wave heating occurred along the spacecraft field line between 15 solar radii and the spacecraft location at the same time that triggered ion acoustic waves were measured at the spacecraft (panel 1A). In panel 1B the wave heating at the beginning of the interval was at least a factor of three greater than the fluctuations of the base temperature through the remainder of the data.

By contrast, the orbit 8 data of Figure 1 indicates significant local wave activity (panel 1G) and many instances of electron heating due to waves at locations in panel 1H when the electron temperature was significantly greater than the red base temperature curve. During each of these times, $T_E/T_I$ of panel 1K was significantly greater than one. By contrast, during about three or four intervals lasting a total of about 20 hours, the measured temperature was close to the base temperature, there was no wave activity at the spacecraft and $T_E/T_I$ was ≤1. The summary of the data in figure 1 is that the base temperature varied by about ±10% while the wave-associated heating was much greater.

Figure 2 presents similar data for PSP perihelia 7 and 9. In both cases, there were significant waves at the spacecraft and significant electron temperature increases associated with large $T_E/T_I$ ratios. Through about four intervals during orbit eight, totaling about 20 hours of low wave activity and small $T_E/T_I$, the measured electron temperature tended to be close to the base temperature. This is the result of little or no wave heating of electrons between 15 solar radii and the measuring point.

Figure 3 gives the electron and ion temperatures versus radial distance, as obtained from averaging the data of orbits 6, 7, 8, and 9. The left panel shows that core electron heating occurred between 20 and 25 solar radii because the core perpendicular and parallel electron temperatures rose above the red base temperature curve between 20 and 25 solar radii and they remained above that curve at larger radial distances. Each plot consists of one-hour running averages of some 10,000,000 raw data points.

Because the ion base temperature at 15 solar radii was strongly dependent on the solar wind speed, it is not possible to determine the ion heating as a function of radial distance from the available data (right panel of figure 3) without further correction of the base ion temperature.

Whistler waves are not seen in the 20-25 solar radius region [Cattell, et al, 2021]. Two classes of electrostatic waves were seen in the current data. One class, represented by their broadband frequency response and short duration, are conventional ion acoustic waves [Mozer et al, 2020]. The other class, represented by their narrowband frequency width and longer durations, are the triggered ion acoustic waves (TIAW) [Mozer et al, 2021] that are associated with the observed electron heating.

Properties of triggered ion acoustic waves are presented in figure 4 and summarized here.
1. Triggered ion acoustic waves are distinguished from normal ion acoustic waves by their narrow band frequency structure, which may be understood [Kellogg, 2021]. Very often, but not always, the triggered ion acoustic waves have the following properties, which are not understood theoretically;
2. A few Hz electric field wave is present, accompanied with bursts of a few hundred to 1000 Hz waves whose bursts are phase locked with each low frequency period (figure 4A).
3. These periodic field structures can exist for large fractions of a day (see Figure 4 of Mozer et al [2021]).
4. Few Hz and few hundred Hz plasma density fluctuations (determined from the spacecraft potential) are present at the two frequencies (figures 4B and 4C).
5. No magnetic field signature is associated with either of these two wave frequencies (figure 4D).
6. The higher frequency electric field and density fluctuations are pure sine waves, indicating that the waves are very narrow band (figures 4E and 4F).

Figures 5A-5H provides expanded views of the two intervals between 21 and 30 solar radii in orbit 7 when the TIAW (and some normal ion acoustic waves) were observed. Looking from perihelion outward in either direction, the red curves in Figure 5C and 5G show the base temperature decrease expected from adiabatic expansion. After perihelion, in panel 5G, there is a significant core, perpendicular electron temperature increase (the black curve in panel 5G) at the time of the TIAW in panel

5E because the temperature further increased above the base temperature level. Before perihelion, in panel 5C, there is a smaller temperature increase at the time of the TIAW. Each of these curves are one-hour running averages made from about 500,000 raw data points.

Mixing of spatial and temporal variations is illustrated in figures 5A and 5C, where heated electrons were also observed at lower altitudes (later times) when there were no TIAW. This situation can result from electrons heated by waves at a lower altitude than the satellite location such that the heated electrons arrived at the spacecraft before or in the absence of the slower moving waves. Thus, one cannot expect a detailed one-to-one correlation between the waves and the heating but there should be a general correlation, such as that shown during all of the heating events.

Data similar to that illustrated in figure 5 are shown for orbit 9 in figure 6. During both the inbound and outbound legs of this orbit, there was significant heating (panels 6C and 6F) at times when $T_E/T_I$ was large (panels 6B and 6E) and TIAW (and some normal ion acoustic waves) were present (panels 6A and 6D).

**Discussion**

The summary, from the statistics available on the altitude range of 15-30 solar radii during four orbits (eight passes through 20-30 solar radii, four of which are shown in figures 5 and 6), is:
1. There were two passes with (Te/Ti)≤1, few waves, and no electron heating.
2. There were six passes with (Te/Ti)>1, with TIAW, and with correlated core electron heating.

The conclusions drawn from this limited data set are;
1. The electron base temperature at 15 solar radii, during about a dozen events lasting about 250 hours, was constant within about ±10%. Thus, the observed temperature, $T_E$, may be modeled as the sum of the base temperature, $T_B$, and the temperature increase, $T_W$, associated with the wave heating between 15 solar radii and the measurement location.
2. The wave heating increased the electron temperature by as much as a factor of two above the base temperature.
3. This core electron heating was associated with triggered ion acoustic waves.

Future PSP orbits will investigate TIAW to provide further statistics on their occurrence and their electron heating. They will also provide higher frequency measurements of plasma density fluctuations (through measurements of the spacecraft potential), which may be a significant contributor to electron heating and to achieving the PSP mission goal of assessing the dominant mechanisms of energy exchange and plasma heating near the Sun.

**Acknowledgements**

This work was supported by NASA contract NNN06AA01C. The authors acknowledge the extraordinary contributions of the Parker Solar Probe spacecraft engineering team at the Applied Physics Laboratory at Johns Hopkins University. The FIELDS experiment on the Parker Solar Probe was designed and developed under NASA contract NNN06AA01C. Our sincere thanks to P. Harvey, K. Goetz, and M. Pulupa for managing the spacecraft commanding, data processing, and data analysis, which has become a heavy load thanks to the complexity of the instruments and the orbit. We also acknowledge the SWEAP team for providing the plasma data. The work of I.V. was supported by NASA Heliophysics Guest Investigator grant 80NSSC21K0581

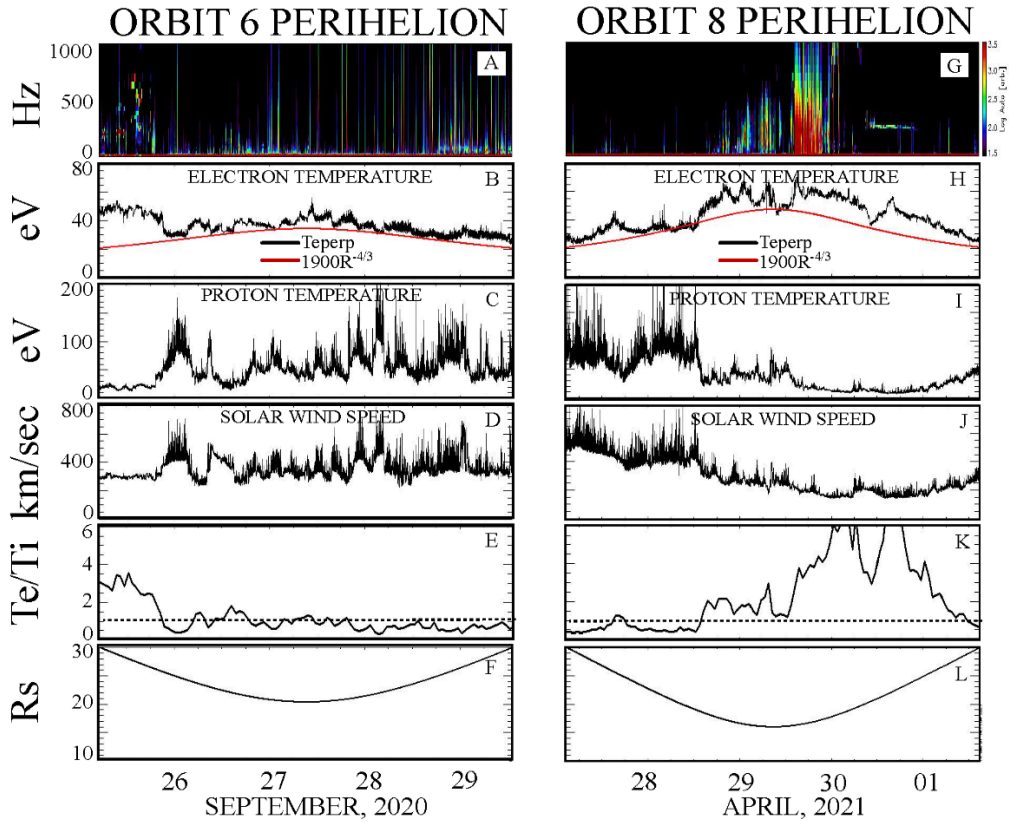

Figure 1. Plasma parameters during perihelion passes six and eight of the Parker Solar Probe. Note that after the first 15% of orbit six, the electron temperature followed the base red curve even though the solar wind speed varied from 300 to 800 km/sec.

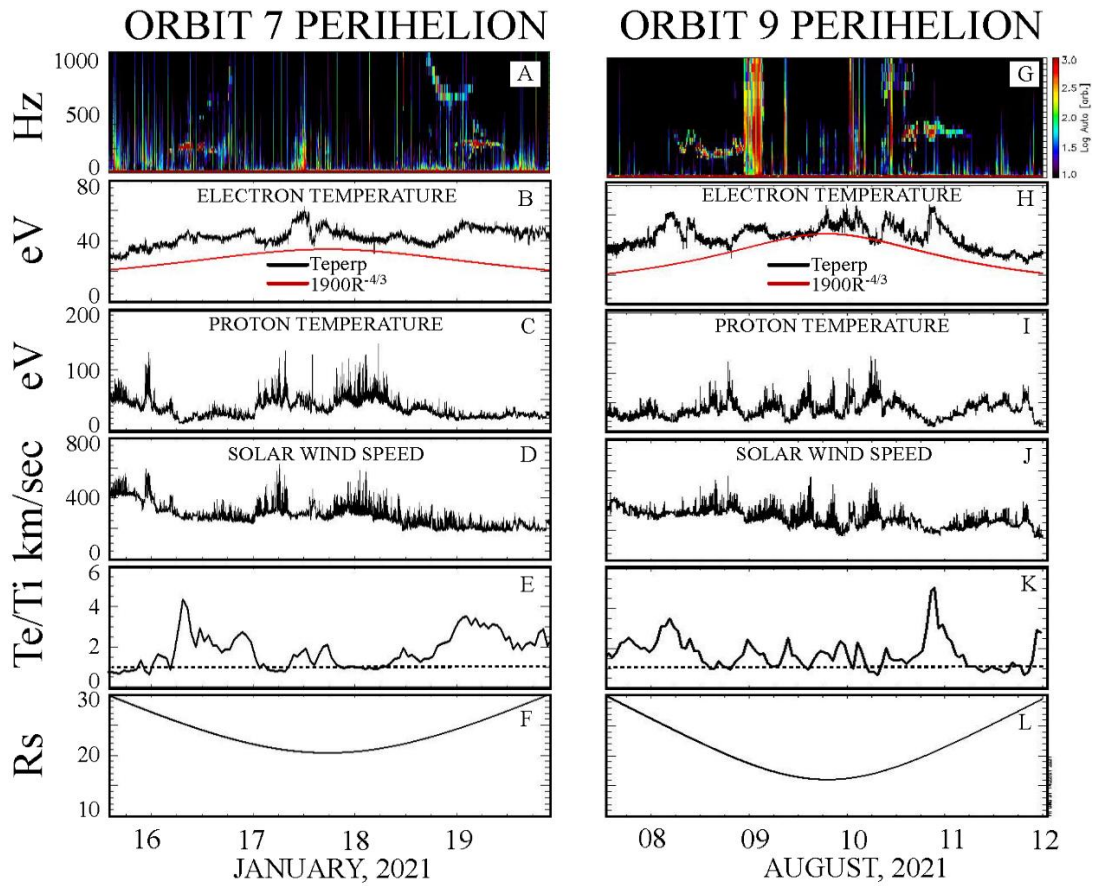

Figure 2. Plasma parameters during perihelion passes seven and nine of the Parker Solar Probe.

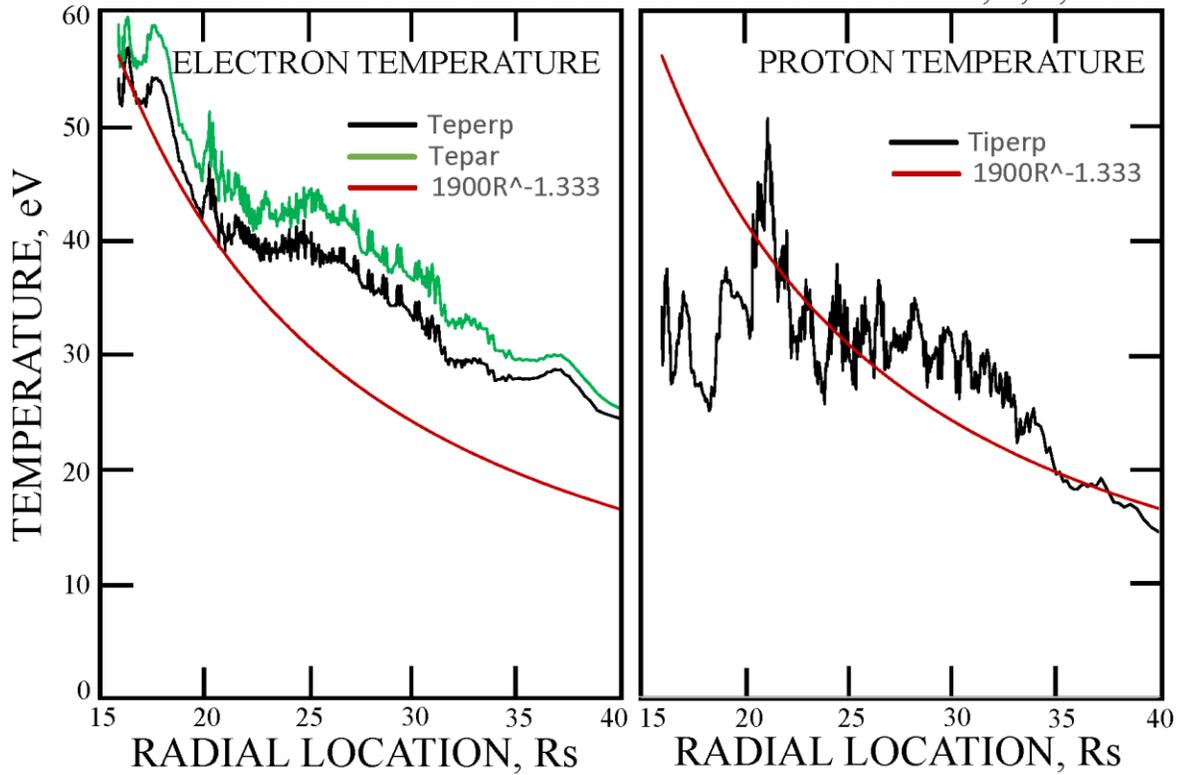

Figure 3. The core electron and ion temperatures versus solar radius, as determined from the eight passes of the Parker Solar Probe through the 15-40 solar radius region on orbits 6, 7, 8, and 9. Each curve gives the one-hour running average of some 10,000,000 raw data points. The red curves give $T_B$, the radial variation of the base temperature for adiabatic expansion with no additional heating. The deviation between the electron core temperatures and the red curve shows that, on the average, electrons were heated in the 20-25 solar radius region. Because the base ion temperature depended much more strongly than did the base electron temperature, it is not possible to determine the ion heating in the right panel without correcting for the large variations in the base ion temperature.

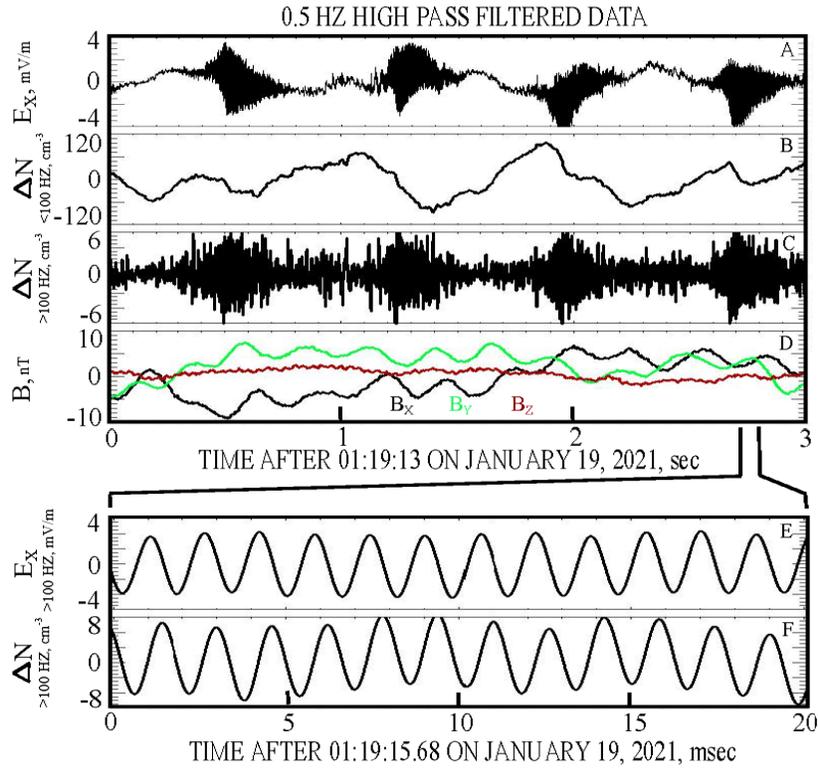
Figure 4. Features of triggered ion acoustic waves, including the electric field having higher frequency pulses in phase with a few Hz wave (Panel 4A), low and high frequency density fluctuations in sync with the electric fields (panels 4B and 4C), the absence of a similar magnetic field signature (panel 4D), and with the higher frequency electric field and density fluctuations being narrow band signals (panels 4E and 4F).

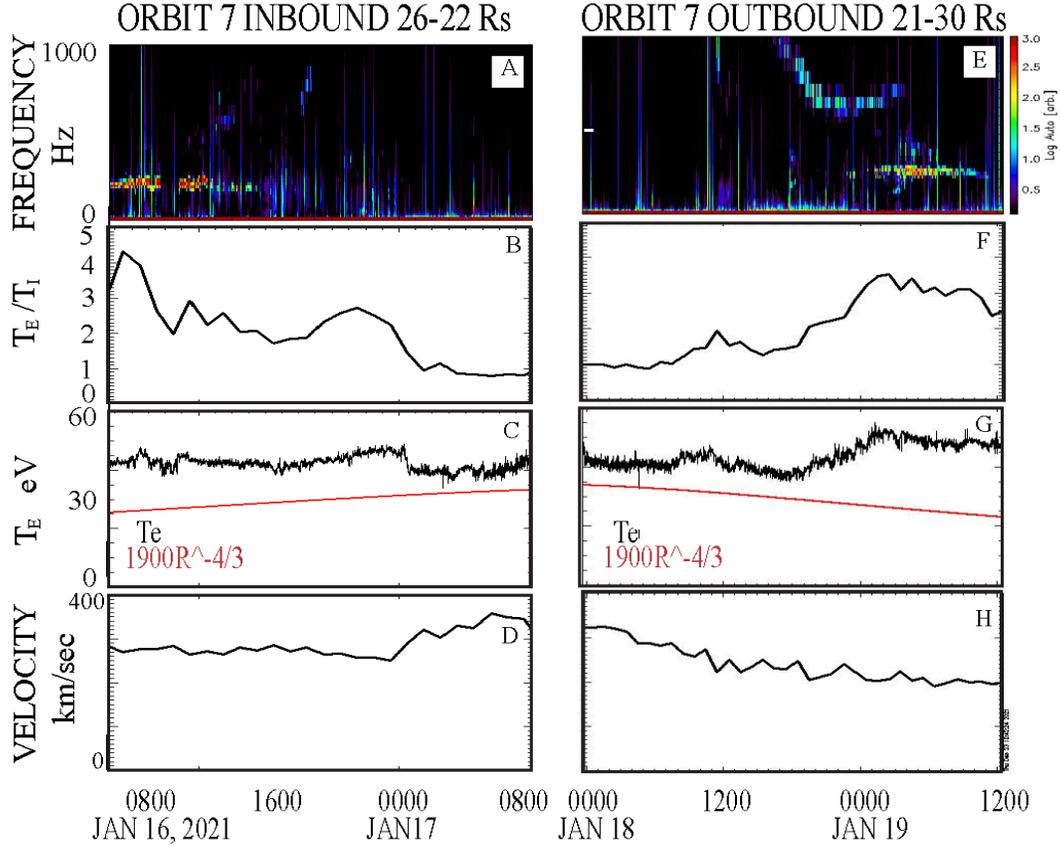

Figure 5. The inbound (panels 5A-5D) and outbound (panels 5E-5H) passes of orbit 7, illustrating the electric field, temperature ratio, electron temperature and the plasma velocity. The red curves in panels 5C and 5G give the base electron temperature decreases expected for adiabatic expansion of the electron gas. Their differences from the measured temperatures (the black curves) show that the electrons were heated in regions containing triggered ion acoustic waves (panels 5A and 5E) and a large electron to ion temperature ratio (panels 5B and 5F).

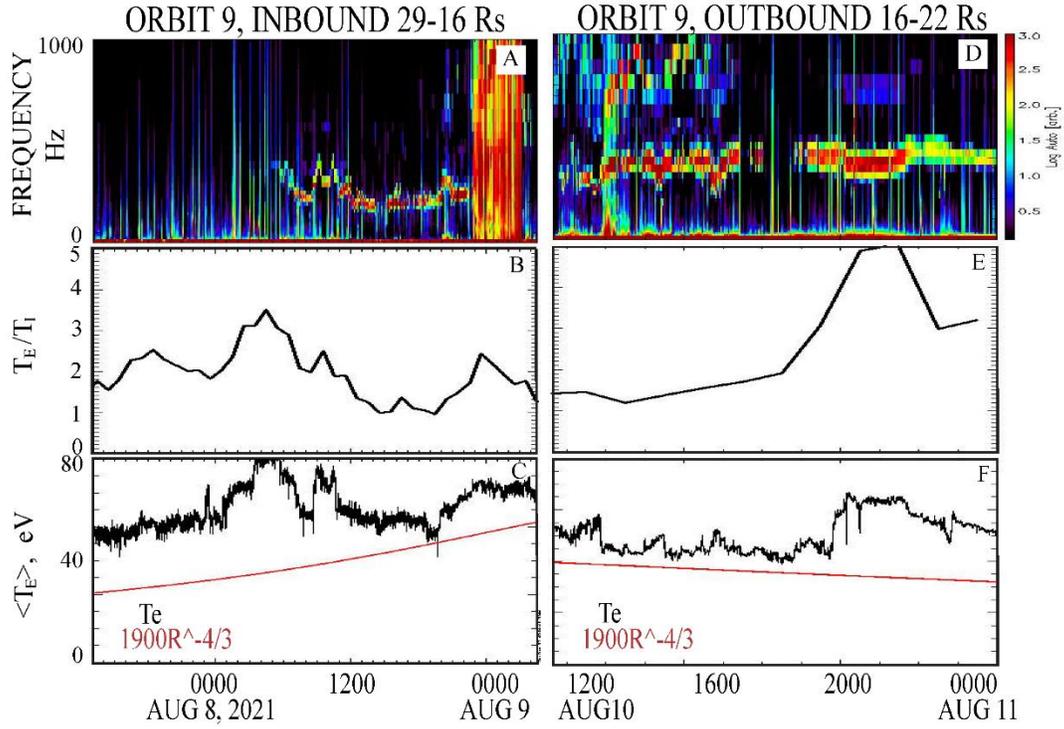

Figure 6. Same as Figure 5 except that this figure is for the perihelion of orbit 9. On both the inbound and outbound passes, triggered ion acoustic waves were present and the electrons were heated in their vicinities.